\documentclass[twocolumn]{aastex62}

\usepackage{color}\usepackage{graphicx}
\usepackage{amsmath}	
\usepackage{amssymb}	
\usepackage{ulem}
\usepackage{chemformula}
\usepackage[figuresleft]{rotating}

\newcommand{\ltsima} {$\; \buildrel < \over \sim \;$}
\newcommand{\gtsima} {$\; \buildrel > \over \sim \;$}
\newcommand{\lta} {\lower.5ex\hbox{\ltsima}}

\newcommand{\gta} {\lower.5ex\hbox{\gtsima}}

\newcommand{\kms}{km\,s$^{-1}$}

\newcommand{\mjybm}{mJy~bm$^{-1}$}

\newcommand{\oiii}{[O\,{\sc iii}]}

\accepted{by ApJ 12-Jun-2023}

\shorttitle{Jet-brightening by circumgalactic molecular gas}
\shortauthors{Emonts et al.}

\begin{document}

\title{CO survey of high-z radio galaxies, revisited with ALMA: Jet-cloud Alignments and Synchrotron Brightening by Molecular Gas in the Circumgalactic Environment}

\correspondingauthor{Bjorn Emonts}
\email{bemonts@nrao.edu}

\author{Bjorn H.\,C. Emonts}
\affiliation{National Radio Astronomy Observatory, 520 Edgemont Road, Charlottesville, VA 22903}

\author{Matthew Lehnert}
\affiliation{Universit\'{e} Lyon 1, ENS de Lyon, CNRS UMR5574, Centre de Recherche Astrophysique de Lyon, F-69230 Saint-Genis-Laval, France}

\author{Sophie Lebowitz}
\affiliation{Steward Observatory, Department of Astronomy, University of Arizona, 993 North Cherry Avenue, Tucson, AZ, 85721, USA}
\affiliation{Department of Astronomy, The Ohio State University, Columbus OH 43210 USA}

\author{George K. Miley}
\affiliation{Leiden Observatory, Leiden University, PO Box 9513, 2300 RA Leiden, The Netherlands}

\author{Montserrat Villar-Mart\'{i}n}
\affiliation{Centro de Astrobiolog\'{i}a, CSIC-INTA, Ctra. de Torrej\'{o}n a Ajalvir, km 4, 28850 Torrej\'{o}n de Ardoz, Madrid, Spain}

\author{Ray Norris}
\affiliation{CSIRO Space \& Astronomy, P.O. Box 76, Epping, NSW 1710, Australia}
\affiliation{School of Science, Western Sydney University, Locked Bag 1797, Penrith, NSW 2751, Australia}

\author{Carlos De Breuck}
\affiliation{European Southern Observatory, Karl Schwarzschild Strasse 2, 85748 Garching bei M$\ddot{u}$nchen Germany}

\author{Chris Carilli}
\affiliation{National Radio Astronomy Observatory, P.O. Box O, Socorro, NM 87801, USA}

\author{Ilana Feain}
\affiliation{Quasar Satellite Technologies PTY Ltd, Cnr Vimiera and Pembroke Roads, Marsfield NSW 2122, Australia}



\begin{abstract}
Powerful radio sources associated with super-massive black holes are among the most luminous objects in the Universe, and are frequently recognized both as cosmological probes and active constituents in the evolution of galaxies. We present alignments between radio jets and cold molecular gas in the environment of distant radio galaxies, and show that the brightness of the radio synchrotron source can be enhanced by its interplay with the molecular gas. Our work is based on CO $J\,>1$ observations with the Atacama Large Millimeter/submillimeter Array (ALMA) of three radio galaxies with redshifts in the range 1.4\,$<$\,$z$\,$<$\,2.1, namely MRC\,0114-211 ($z$\,=\,1.41), MRC\,0156-252 ($z$\,=\,2.02), and MRC\,2048-272 ($z$\,=\,2.05). These ALMA observations support previous work that found molecular gas out to 50 kpc in the circumgalactic environment, based on a CO(1-0) survey performed with the Australia Telescope Compact Array (ATCA). The CO emission is found along the radio axes but beyond the main radio lobes. When compared to a large sample of high-$z$ radio galaxies from the literature, we find that the presence of this cold molecular medium correlates with an increased flux-density ratio of the main vs. counter lobe. This suggest that the radio lobe brightens when encountering cold molecular gas in the environment. While part of the molecular gas is likely related to the interstellar medium (ISM) from either the host or a companion galaxy, a significant fraction of the molecular gas in these systems shows very low excitation, with r$_{2-1/1-0}$ and r$_{3-2/1-0}$ values $\lesssim$0.2. This could be part of the circumgalactic medium (CGM). 
\end{abstract}



\section{Introduction}
\label{sec:intro}

High-$z$ radio galaxies are among the most massive and active systems in the Early Universe \citep{mil08}. They are key objects for better understanding a range of astrophysical phenomena, from black-hole activity to galaxy and cluster formation \citep[e.g.,][]{ver01,pen01,sma03,ste03,kur03,roc04,ove06,kod07,sey07,vil07a,bre10,bar12,gal12,wyl13,dan14,fal19}. High-$z$ radio galaxies at $z$\,$>$\,1 are typically defined as having a radio luminosity at 500 MHz in the restframe of $L_{\rm 500}\,>\,10^{27.5}$ W\,Hz$^{-1}$ \citep{mil08}. Their bright radio sources originate from the accretion-disk region around a supermassive black hole, where biconical magnetic fields accelerate charged particles that emit synchrotron radiation (see reviews by \citealt{mil80} and \citealt{bla01}). This process creates radio jets and lobes, which can propagate well beyond the host galaxy and are often hundreds of kpc in size \citep[e.g.,][]{car97,pen00,bar00}.

Because their bright radio continuum is easily detectable, high-$z$ radio galaxies have long acted as beacons for tracing distant galaxies \citep[e.g.,][]{rot94,rot97,bre00,sax18,bro22}. Therefore, high-$z$ radio galaxies were among the first laboratories for studying galaxy formation in the Early Universe \citep[e.g.,][]{spi81,mcc87lya,mcc90,cha90}. Deep follow-up optical and infra-red imaging and spectroscopy showed that their hosts are massive galaxies at redshifts of $z$\,$\sim$\,2$-$5 \citep[see review by \citealt{mcc93}; also][]{djo88,leh92,mcc95,cha96a,cha96b,ste99,bre01}. In addition to being beacons for tracing distant galaxies, the powerful high-$z$ radio sources also play a key role in the formation and evolution of galaxies and galaxy clusters, for example by driving massive outflows of gas \citep{nes17}. Intriguing `alignment effects' have also been observed between the radio jets and various constituents of the radio host galaxy \citep{mil08}. This includes alignments with the ultra-violet (UV) rest-frame continuum and submillimeter emission, which has been interpreted as jet-triggered star formation \citep{cha87,mcc87,beg89,you89,ree89,pen99,bic00,ste03,ivi12}.

High-$z$ radio galaxies were also among the first objects observed to contain a rich circumgalactic medium (CGM), primarily detected in the form of giant Ly$\alpha$ nebulae of ionized gas \citep{mcc87lya,cha90,hec91,oji96,vil02,vil03,vil06,vil07b,reu03,reu07,mil06,hum08,swi15,ver17,fal21}. These Ly$\alpha$ nebulae often show kinematically perturbed gas along the radio axis, with more quiescent gaseous halos detected beyond the extent of radio source \citep{vil03,hum06}. Ionized gas nebulae associated with high-$z$ radio galaxies are also seen in H$\alpha$ and are often enriched with heavy elements across tens of kpc \citep[e.g.,][]{mcc92,car01,kur02,vil03,swi15,nes17,fal21}. In some cases, cold molecular gas is found in the halo environment of high-$z$ radio galaxies \citep{kla04,kla05,ivi12,emo14,emo15a,emo16,gul16a,fal21,li21,wan21,bre22}.

\citet{kla04} discovered that CO emission from cold molecular gas is often offset from the host galaxy and preferentially aligned along the radio axis. Other studies showed similar alignments between the radio source and cold molecular halo gas, including radio-loud quasars at $z$\,$\sim$\,0.3 \citep{ara11,pap08,elb09} and $z$\,$\sim$\,2.2 \citep{li21}, as well as the radio galaxies TXS\,0828+193 at $z$\,$\sim$\,2.6 \citep{nes09,fog21} and TN\,J0924-2201 at $z$\,$\sim$\,5.2 \citep{lee23}. This alignment effect was also revealed by a CO(1-0) survey performed with the Australia Telescope Compact Array (ATCA) \citep[][hereafter Paper I]{emo14}. 

The ATCA survey from Paper\,I utilized the ultra-compact array configurations of the ATCA to obtain exquisite surface-brightness sensitivity for detecting CO(1-0) emission on scales of tens of kpc in a representative sample of high-$z$ radio galaxies from the flux-limited 408 MHz Molonglo Reference (MRC) Catalogue \citep{lar81}. This ATCA survey revealed that five out of 13 high-$z$ radio galaxies in the sample (38$\%$) contain detectable amounts of CO(1-0) in the halo environment, preferentially aligned along the radio axis. One of these systems, MRC\,0152-209 (Dragonfly galaxy), was observed with ALMA at high resolution in CO(6-5), which revealed that the radio jet aligns and interacts with molecular gas in the disk of a merging companion galaxy \citep{leb23}. For three other systems observed as part of the ATCA survey, namely MRC\,0114-211, MRC\,0156-252, and MRC\,2048-272, all the CO(1-0) emission was detected in a single molecular gas reservoir, located beyond the brightest edge of the radio source. As shown in Paper I, even though the CO(1-0) luminosity of this halo gas is similar to what is typically found in submillimeter galaxies (SMGs), the peaks of the CO(1-0) emission coincide with regions that are devoid of 4.5$\mu$m infra-red emission down to $1-2$ margnitudes below $L^{*}$, based on observations with the Infrared Array Camera (IRAC) Band 2 on the {\it Spitzer} Space Telescope \citep{gal12,wyl13}.

This paper further investigates the molecular gas in the three high-$z$ radio galaxies MRC\,0114-211, MRC\,0156-252, and MRC\,2048-272, through CO(2-1) and CO(3-2) observations performed with the Atacama Large Millimeter/submillimeter Array (ALMA). A comparison of the properties and spatial distributions between the CO-emitting gas and the radio synchrotron emission will be used to further study the observed alignments between radio sources and cold gas. The goal is to understand the nature of interaction between the radio source and molecular gas in the halos of high-$z$ radio galaxies.

Throughout this paper, we shall assume the same cosmological parameters as in Paper I, namely H$_{0}$\,=\,71 km\,s$^{-1}$\,Mpc$^{-1}$, $\Omega_{\rm M}$\,=\,0.27, and $\Omega_{\rm \lambda}$\,=\,0.73 \citep{wri06}.

\section{ALMA data}
\label{sec:observations}

MRC\,0114-211 ($z$\,=\,1.41), MRC\,0156-252 ($z$\,=\,2.02), and MRC\,2048-272 ($z$\,=\,2.05) were observed with the 12m Atacama Large Millimeter Array (ALMA) and 7m Atacama Compact Array (ACA) as part of ALMA Cycle 3 (ID: 2015.1.00897.S). We targeted the lowest transition of CO that is observable with ALMA, which for MRC\,0114-211 is CO(2-1) and for MRC\,0156-252 and MRC\,2048-272 is CO(3-2), all covered in Band 3 \citep{cla05}. The ALMA 12m array observed MRC~0114-211, MRC~0156-252, and MRC 2048-272 during 12$-$19 Jan 2016 in C36-1 configuration for 56, 138, and 111 minutes, respectively, which included time for slewing and calibration. The 7m ACA observations were observing during Jan$-$Sept 2016, with corresponding total observing times of 3.7, 9.4, and 7.6 hours, not including observations that failed ALMA's quality assurance process. We used four spectral windows of 1.875 GHz with 2 MHz channels. For MRC\,0114-211 and MRC\,0156-211, one of the spectral windows was centred around the redshifted CO line, while the remaining three covered line-free continuum emission. For MRC\,2048-272, which had a wider overall CO(1-0) signal (Paper\,I), two of the four spectral windows were stitched together to cover the CO line, while the other two observed only continuum emission. A standard strategy for bandpass, phase, and flux calibration was adopted as part of the ALMA calibration plan.

The ALMA data were reduced with the Common Astronomy Software Applications (CASA; \citealt{casa22}). Calibration was performed with the scriptForPI.py calibration scripts that were included with the archival data set, using CASA v.4.5.1 with pipeline v.r35932 for the 12m data of MRC~0114-211 and MRC~2048-272, CASA v.4.5.3 with pipeline r36115 for the 7m data of MRC~0114-211, CASA v.4.6.0 (without pipeline) for the 12m data of MRC~0156-252, and CASA v.4.7.0 with pipeline v.r38377 for the 7m data of MRC~0156-252 and MRC~2048-272. 

Images were made using CASA v.5.6.0. First, we subtracted the continuum emission using CASA task {\sc uvcontsub} by fitting a straight line to the line-free channels for each visibility. We then imaged the continuum-subtracted line emission using CASA task {\sc tclean}, utilizing the {\sc mosaic} gridder to combine the 12m and 7m data. We also performed a correction for the primary beam response. After binning the image cube to a channel width that best visualizes the line emission, we applied a Hanning smooth to optimize the CO signals. This Hanning smooth created an effective velocity resolution that is twice the channel width. Because of the faintness of the CO, no deconvolution was applied (but see Appendix \ref{sec:companion}). Table \ref{tab:data} summarizes the spatial resolution (synthesized beam size), astrometric accuracy, spectral resolution, and root-mean-square (rms) noise level of the image products after combining the 12m and 7m data.

\begin{figure*}
\centering
\includegraphics[width=0.95\textwidth]{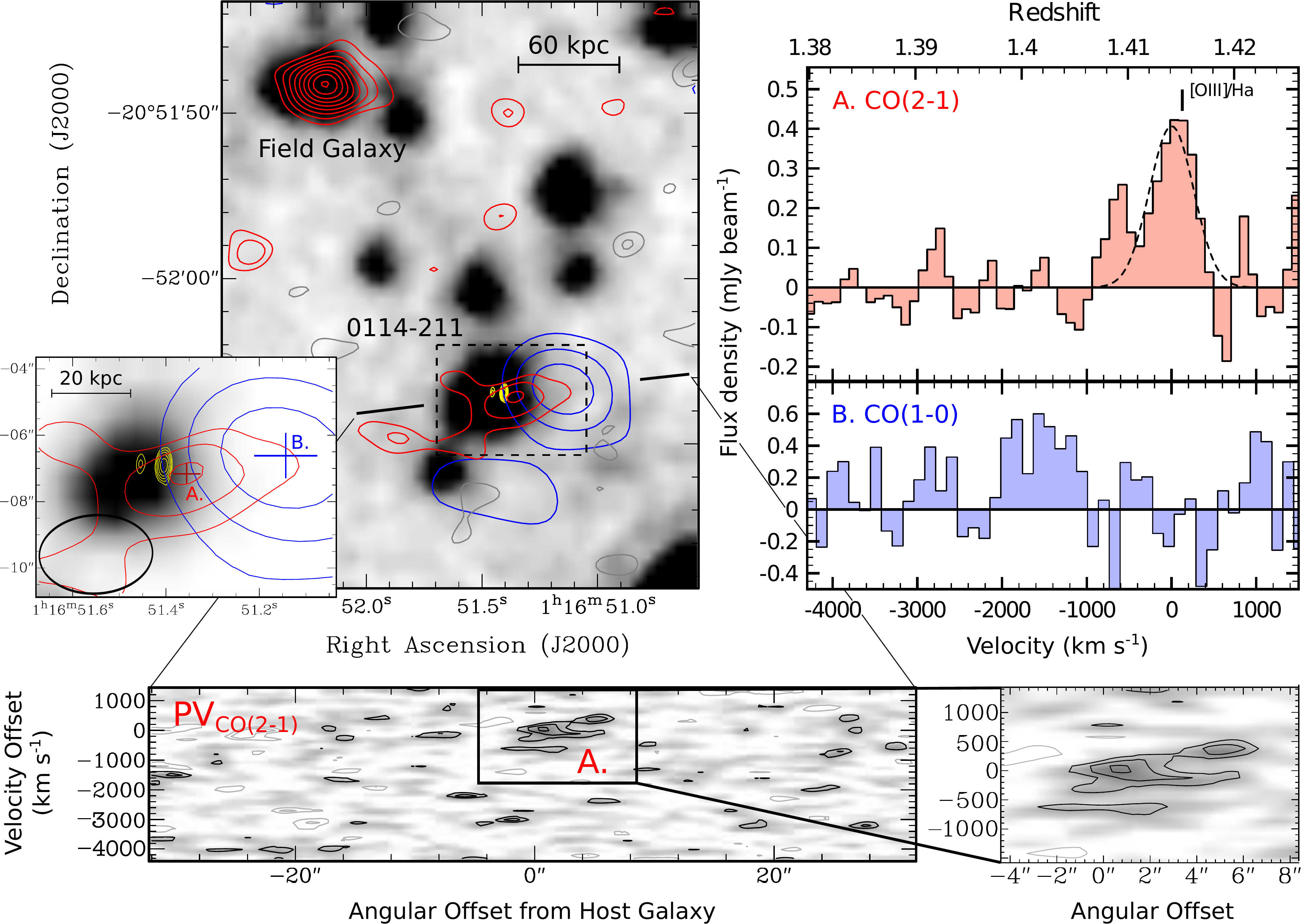}
\caption{Molecular gas in the environment of MRC~0114-211 ($z$\,=\,1.41). Top-left: IRAC 4.5$\mu$m image of the environment of MRC~0114-211 \citep{gal12,wyl13}, with overlaid in red the contours of CO(2-1) observed with ALMA across the velocity range -773 to 227 \kms\ and in blue the contours of CO(1-0) observed with ATCA (see Paper\,I). CO contour levels start at 2$\sigma$ and increase with 1$\sigma$, with $\sigma$\,=\,0.067 Jy\,bm$^{-1}$\,$\times$\,\kms\ for CO(2-1) and $\sigma$\,=\,0.094 Jy\,bm$^{-1}$\,$\times$\,\kms\ for CO(1-0). Negative contours are shown in grey for CO(2-1), while for clarity they are omitted for CO(1-0). The yellow contours show a Karl G. Jansky Very Large Array (VLA) 8.2\,GHz image of the radio synchrotron source \citep{bre10}, with contour levels starting at 3 mJy\,beam$^{-1}$ and increasing by a factor of 3. The panel on the bottom-left corner shows a zoom-in of the region around MRC\,0114-211. The crosses mark the locations of the CO emission-line peaks, while their sizes indicate the astrometric uncertainties (see Table\,\ref{tab:data}). The ALMA beam is indicated with the black ellipse in the bottom-left corner. Top-right: Spectra of CO(2-1) (top - red) and CO(1-0) (bottom - blue) associated with MRC~0114-211. The spectra were taken at the locations of the peaks of the CO emission, which are marked with crosses in the total intensity image shown in the left zoom-in panel. The dashed line in the top plot shows a single Gaussian fit to the CO(2-1) profile, which was used to determine the redshift $z_{\rm CO(2-1)}$\,=\,1.414\,$\pm$\,0.001. The vertical bar shows the redshift determined from [OIII] and H$\alpha$ \citep{nes17}. The spectrum of CO(1-0) from Paper\,I was shifted in velocity to match the redshift of the CO(2-1) line. Bottom: Position-Velocity (PV) diagram of the ALMA CO(2-1) emission taken along a 1-dimensional line in the direction that intersects the peak emission in both CO(2-1) and CO(1-0), as indicated in the top-left plot. Contour levels are at 2, 3, 4$\sigma$, with $\sigma$ the rms noise level from Table \ref{tab:data} (negative contours are in grey). The PV plot was not corrected for the primary beam response in order to visualize the noise level. The right panel shows a zoom-in on the CO(2-1) emission.} 
\label{fig:mrc0114}
\end{figure*}

\begin{figure*}
\centering
\includegraphics[width=\textwidth]{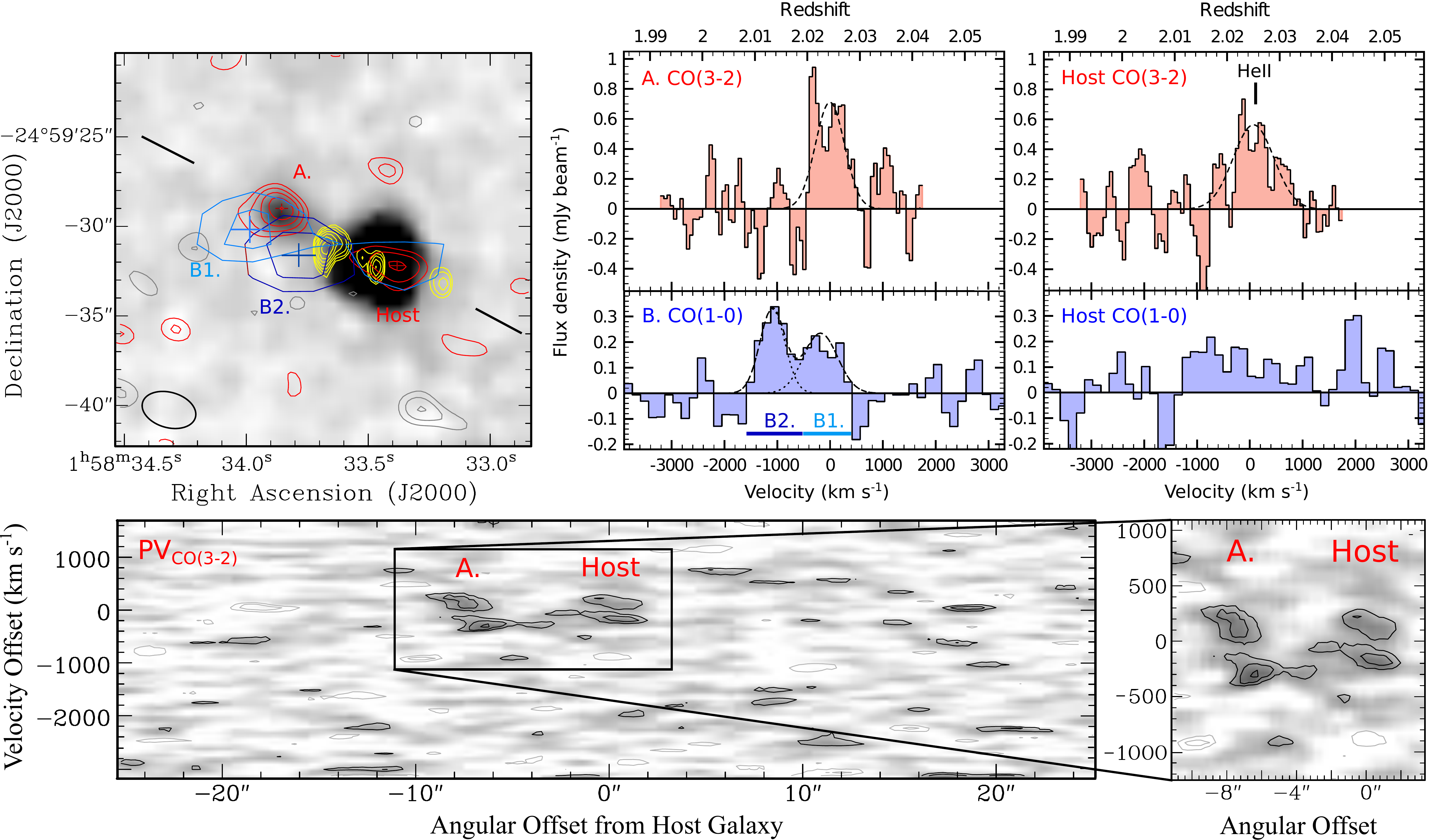}
\caption{Molecular gas in the environment of MRC~0156-252 ($z$\,=\,2.02). Top-left: IRAC 4.5$\mu$m image of the environment of MRC~0156-252 \citep{gal12,wyl13}, with overlaid in red the contours of CO(3-2) observed with ALMA across the velocity range -417 to 363 \kms\ and in blue the contours of CO(1-0) observed with ATCA (see Paper\,I). The CO(1-0) emission found with ATCA is split into two components, with component B1 covering the velocity range -1583 to -508 \kms\ and component B2 covering -508 to 417 \kms\ (see Fig. 7 of Paper\,I, corrected for the redshift of CO(3-2)). CO contour levels start at 2$\sigma$ and increase with 1$\sigma$, with $\sigma$\,=\,0.080 Jy\,bm$^{-1}$\,$\times$\,\kms\ for CO(3-2), and $\sigma$\,=\,0.055 and 0.064 Jy\,bm$^{-1}$\,$\times$\,\kms\ for CO(1-0) components B1 and B2, respectively. Negative contours are shown in grey for CO(3-2), while for clarity they are omitted for CO(1-0). The crosses mark the locations of the CO emission-line peaks, while their sizes indicate the astrometric uncertainties (see Table\,\ref{tab:data}). The ALMA beam is indicated with the black ellipse in the bottom-left corner. The yellow contours show a 4.7 GHz VLA image of the radio synchrotron source from \citet{car97}, with contour levels starting at 0.58 mJy\,beam$^{-1}$ and increasing by a factor of 2. Top-right: Spectra of CO(3-2) (top - red) and CO(1-0) (bottom - blue) associated with the regions `A/B' and `Host'. The CO(3-2) spectrum of region `A', as well as the CO(3-2) and CO(1-0) spectra of region `Host', were taken at the locations of the peaks of the CO(3-2) emission detected with ALMA, which are marked with small red crosses in the total intensity image shown in the left panel. The CO(1-0) spectrum in region `A/B' was taken at the peak of the overall CO(1-0) emission from Paper\,I, taking into consideration that the ATCA beam has a large size of 6.9$^{\prime\prime}$\,$\times$\,4.9$^{\prime\prime}$ (71$^{\circ}$). The dashed lines show Gaussian fits to the profiles. A single Gaussian fit to the CO(3-2) profile was used to determine the redshift $z_{\rm CO(3-2)}$\,=\,2.024\,$\pm$\,0.001. The vertical bar shows the redshift from \ion{He}{2} derived by \citet{gal13}. The spectrum of CO(1-0) from Paper\,I was shifted in velocity to match the redshift of the CO(3-2) line. Bottom: Position-Velocity (PV) diagram of the ALMA CO(3-2) emission taken along a 1-dimensional line in the direction that intersects the peak emission in both region A and the core, as indicated in the top-left plot. Contour levels are at 2, 3, 4$\sigma$, with $\sigma$ the rms noise level from Table \ref{tab:data} (negative contours are in grey). The PV plot was not corrected for the primary beam response in order to visualize the noise level. The right panel shows a zoom-in on the CO(3-2) emission.} 
\label{fig:mrc0156}
\end{figure*}

\begin{deluxetable*}{ccccccc}[htb!]
  \tablecaption{ALMA CO Line Data Properties}
\tablecolumns{7}
\tablenum{1}
\tablewidth{0pt}
\tablehead{
\colhead{Source} & {CO($J$,$J-1$)} & {Beamsize (PA)} & $\Delta$Astrometry$^{\dagger}$ & {Channel width} & {Spectral resolution} & {rms noise$^{\ddagger}$} \\
 {MRC} & {transition} & {(arcsec$^{2}$)} & (arcsec) & {(\kms)} & {(\kms)} & {(mJy\,bm$^{-1}$\,chan$^{-1}$)}
}
\startdata
0114-211 & $J$=2  & 3.47\,$\times$\,2.43 (-86.8$^{\circ}$) & 0.4 & 100  &  200 & 0.11 \\
0156-252 & $J$=3 & 3.04\,$\times$\,2.00 (74.1$^{\circ}$) & 0.3 & 60 & 120 & 0.21 \\
2048-262 & $J$=3 & 3.35\,$\times$\,1.98 (87.0$^{\circ}$) & 0.4 & 100 & 200 & 0.14 \\
\enddata
\tablecomments{$^{\dagger}$ The astrometric uncertainty of the CO detections discussed in this paper is dominated by the limited signal-to-noise (S/N), which results in small positional errors with respect to the phase center. This astrometric error is calculated using $\delta \theta_{\rm rms}$\,=\,$\frac{1}{2}$ $\langle \Theta_{\rm beam} \rangle$\,(S/N)$^{-1}$ \citep{pap08}. For values in this table, $\Delta$Astrometry\,$\equiv$\,$\delta \theta_{\rm rms}$, and $\langle \Theta_{\rm beam} \rangle$ is the major axis of the beamsize. An additional uncertainty in the absolute astrometry, due to phase errors introduced by errors in the baseline length, is calculated using $\delta \theta_{\rm bas}$\,=\,($\delta${\textbf {\textsl B}}\,$\cdot$\,$\Delta${\textbf {\textsl k}})/{\textbf {\textsl B}}\,$\approx$\,($\delta \Phi_{\rm bas}$/2$\pi$)$\langle \Theta_{\rm beam} \rangle$ \citep{pap08}, with $\delta \Phi_{\rm bas
}$\,$\sim$\,(2$\pi$/$\lambda$)($\delta {\textbf {\textsl B}}$\,$\cdot$\,$\Delta {\textbf {\textsl k}}$), $\Delta {\textbf {\textsl k}}$ the distance to the phase calibrator, and $|\delta {\textbf {\textsl B}}|$ the calibration error of the baseline length. For ALMA, $|\delta {\textbf {\textsl B}}|$\,$\sim$\,0.2\,mm\,km$^{-1}$ \citep{hun16}, with maximum 12m baselines in our data of $\la$350m, while for ATCA we assume $|\delta {\textbf {\textsl B}}|$\,$\sim$\,1\,mm. This means that $\Delta \theta_{\rm bas}$\,$<<$\,$\delta \theta_{\rm rms}$ for the data presented in this paper.}
\tablecomments{$^{\ddagger}$ rms noise before primary beam correction and after Hanning smoothing.}
\label{tab:data}
\end{deluxetable*}

\begin{figure*}
\centering
\includegraphics[width=0.95\textwidth]{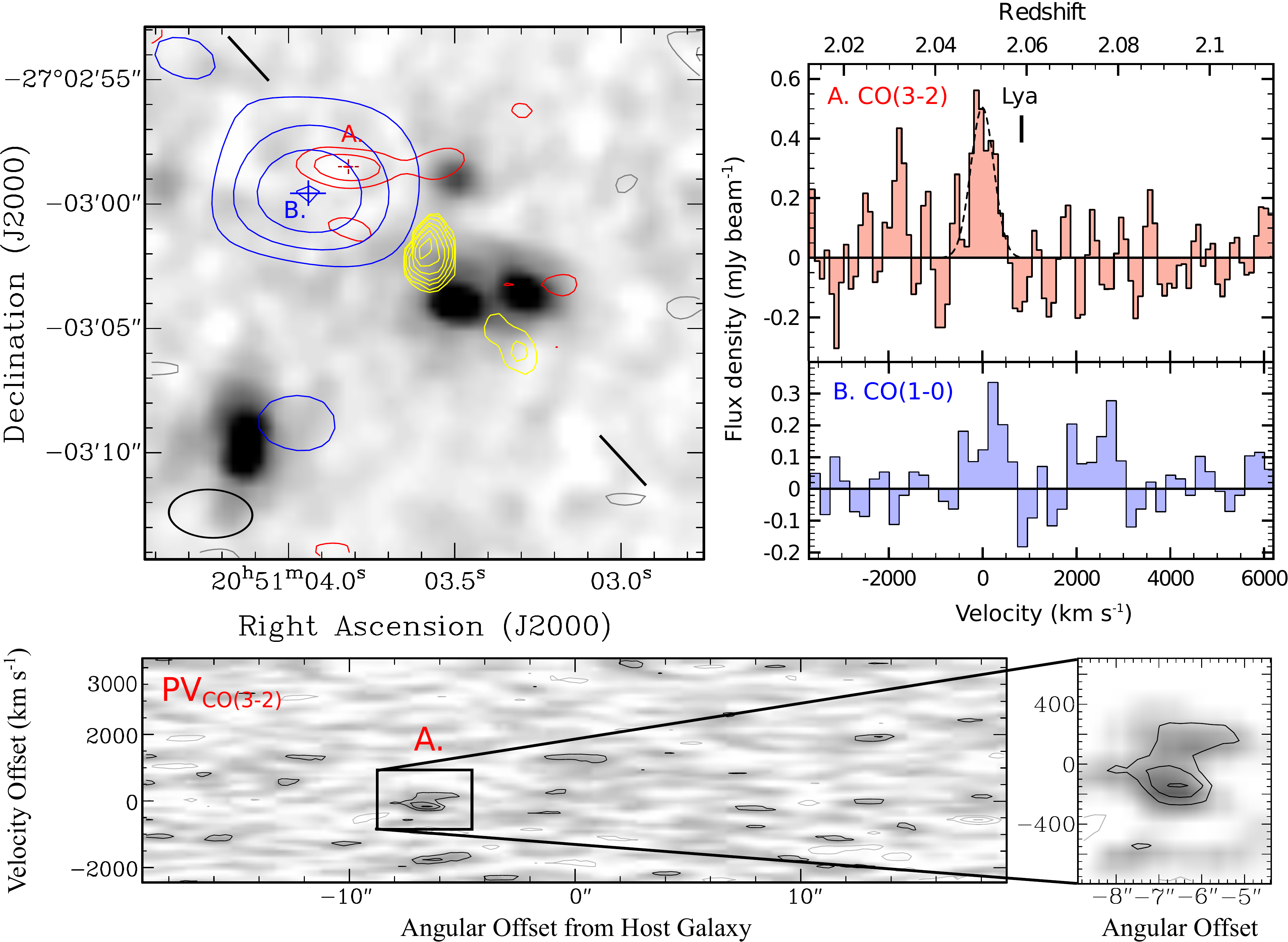}
\caption{Molecular gas in the environment of MRC~2048-272 ($z$\,=\,2.05). Left: IRAC 4.5$\mu$m image of the environment of MRC~2048-272 \citep{gal12,wyl13}, with overlaid in red the contours of a tentative CO(3-2) detection observed with ALMA across the velocity range -265 to 235 \kms\ and in blue the contours of CO(1-0) observed with ATCA \citep[see][]{emo14}. CO contour levels start at 2$\sigma$ and increase with 1$\sigma$, with $\sigma$\,=\,0.056 Jy\,bm$^{-1}$\,$\times$\,\kms\ for CO(3-2) and $\sigma$\,=\,0.080 Jy\,bm$^{-1}$\,$\times$\,\kms\ for CO(1-0). Negative contours are shown in grey for CO(3-2), while for clarity they are omitted for CO(1-0). The crosses mark the location of the CO emission-line peaks, while their sizes indicate the astrometric uncertainties (see Table\,\ref{tab:data}). The ALMA beam is indicated with the black ellipse in the bottom-left corner. The yellow contours show a 4.7 GHz VLA image of the radio synchrotron source from \citet{pen00}, with contour levels starting at 0.5 mJy\,beam$^{-1}$ and increasing by a factor of 2. Right: Spectra of CO(3-2) (top - red) and CO(1-0) (bottom - blue) associated with MRC~2048-272. The spectra were taken at the locations of the peaks of the CO emission, which are marked with crosses in the total intensity image shown in the left panel. The dashed line in the top plot shows a single Gaussian fit to the CO(3-2) profile, which was used to determine the redshift $z_{\rm CO(3-2)}$\,=\,2.050\,$\pm$\,0.001. The vertical bar shows the redshift from  Ly$\alpha$ derived by \citet{ven07}. The spectrum of CO(1-0) from Paper\,I was shifted in velocity to match the redshift of the CO(3-2) line. Bottom: Position-Velocity (PV) diagram of the ALMA CO(3-2) emission taken along a 1-dimensional line along the radio axis, which also intersects the peak emission of CO(3-2). Contour levels are at 2, 3, 4$\sigma$, with $\sigma$ the rms noise level from Table \ref{tab:data} (negative contours are in grey). The PV plot was not corrected for the primary beam response in order to visualize the noise level. The right panel shows a zoom-in on the CO(3-2) emission.} 
\label{fig:mrc2048}
\end{figure*}

\begin{deluxetable*}{llcccccc}[htb!]
  \tablecaption{ALMA CO Detections}
\tablecolumns{8}
\tablenum{2}
\tablewidth{0pt}
\tablehead{
\colhead{MRC Source} & Region & {CO($J$,$J-1$)} & {$z_{{\rm CO}(J,J-1)}$} & {FWHM$_{{\rm CO}(J,J-1)}$} & {$I_{{\rm CO}(J,J-1)}$}$^{*}$ & {$L^{\prime}_{{\rm CO}(J,J-1)}$} & {r$_{(J,J-1)/1-0}$}$^{\dagger}$ \\
     & & {transition} & & {(km/s)} &  & {(K\,km\,s$^{-1}$\,pc$^{2}$)} & 
}
\startdata
0114-211 (A)$^{\ddagger}$  & ISM/CGM & $J$=2 & 1.414\,$\pm$\,0.001 & 520\,$\pm$\,120 & 0.23\,$\pm$\,0.04 & (6.1\,$\pm$\,1.1)\,$\times$\,10$^{9}$ & $>$0.34 \\
0156-252 (A/B1)$^{\S}$ & Companion & $J$=3 & 2.024\,$\pm$\,0.001 & 655\,$\pm$\,80 & 0.51\,$\pm$\,0.07 & (1.2\,$\pm$\,0.2)\,$\times$\,10$^{10}$ & 0.30\,$\pm$\,0.08  \\
0156-252 (C) & Host & $J$=3 & 2.025\,$\pm$\,0.001  & 910\,$\pm$\,135 & 0.55\,$\pm$\,0.09 &  (1.3\,$\pm$\,0.2)\,$\times$\,10$^{10}$ & $>$0.32 \\
2048-262 (A) & CGM-1 & $J$=3 & 2.050\,$\pm$\,0.001 & 540\,$\pm$\,95 & 0.29\,$\pm$\,0.05 &  (6.8\,$\pm$\,1.1)\,$\times$\,10$^{9}$ & 0.15\,$\pm$\,0.04 \\
\enddata
\tablecomments{$^{*}$ Units for $I_{{\rm CO}(J,J-1)}$ are in Jy\,bm$^{-1}$\,$\times$\,km\,s$^{-1}$. Uncertainties include the uncertainty in the Gaussian fitting, plus an assumed 5$\%$ uncertainty in absolute flux calibration of the ALMA data.}
\tablecomments{$^{\dagger}$The value of r$_{J,J-1/1-0}$ is the line ratio of the CO luminosities $L^{\prime}_{\rm CO(J,J-1)}$ between the ALMA ($J$=3/2) and ATCA ($J$=1) data from Paper I. This value reflects the excitation conditions of the molecular gas.}
\tablecomments{$^{\ddagger}$For MRC\,0114-211, no CO(1-0) is detected in region A in the velocity range of the CO(2-1) detection. We assume a 3$\sigma$ upper limit across the FWHM of the CO(2-1) detection.}
\tablecomments{$^{\S}$For MRC\,0156-252, the CO(3-2) in region A and CO(1-0) in region B1 occur at the same redshift. Because the spatial shift between the CO peaks in these regions is less than half the FWHM of the synthesized ATCA beam, we obtained the line ratio by comparing the peak CO(3-2) emission in region A with the peak CO(1-0) emission in region B1, despite the fact that they are not exactly co-spatial in Fig.\,\ref{fig:mrc0156}. The values for the Gaussian fitting of the CO(1-0) component of region B1 [B2] are: FWHM$_{\rm CO(1-0)}$\,=\,755\,$\pm$\,165 [555\,$\pm$\,110] \kms\ and $I_{\rm CO(1-0)}$\,=\,0.19\,$\pm$\,0.04 [0.20\,$\pm$\,0.04] mJy\,bm$^{-1}$\,$\times$\,km\,s$^{-1}$.}
\label{tab:detect}
\end{deluxetable*}

\begin{deluxetable*}{llcccc}[htb!]
  \tablecaption{ALMA CO Non-detections in CGM (after smoothing)}
\tablecolumns{6}
\tablenum{3}
\tablewidth{0pt}
\tablehead{
\colhead{Source} & Region & rms (smoothed) & {$I_{{\rm CO}(J,J-1)}$}$^{\dagger}$ & {$I_{\rm CO(1-0)}$}$^{\ddagger}$ & {r$_{(J,J-1)/1-0}$} \\
    {MRC} & & (mJy\,bm$^{-1}$\,chan$^{-1}_{\rm 100}$) & {(Jy\,bm$^{-1}$\,$\times$\,km\,s$^{-1}$)} & {(Jy\,bm$^{-1}$\,$\times$\,km\,s$^{-1}$)} & 
}
\startdata
0114-211 (B)  & CGM & 0.32 & $<$0.23 & 0.43\,$\pm$\,0.08 & $<$0.13 \\
0156-252 (B2) & ISM/CGM & 0.55 & $<$0.39 & 0.20\,$\pm$\,0.04 &  $<$0.22 \\
2048-262 (B) & CGM-2 & 0.35 & $<$0.26 & 0.19\,$\pm$\,0.05 & $<$0.16 \\
\enddata
\tablecomments{$^{*}$Rms noise level per 100 km\,s$^{-1}$ channel, after tapering and smoothing the ALMA data to the same resolution as the ATCA CO(1-0) data from Paper I.}
\tablecomments{$^{\dagger}$For the ALMA non-detections in the CGM, we use conservative limits based on the tapered/smoothing ALMA data and assuming a 3$\sigma$ limit over the FWHM of the CO(1-0) emission.}
\tablecomments{$^{\ddagger}$See Paper\,I and caption of Table \ref{tab:detect} for details on the CO(1-0) detections, with assumption FWHM\,=\,0.5\,$\times$\,FWZI.}
\label{tab:nondetect}
\end{deluxetable*}

\section{results}

\subsection{ALMA results}
\label{sec:almaresults}

Figures \ref{fig:mrc0114} - \ref{fig:mrc2048} show the ALMA imaging of MRC\,0114-211, MRC\,0156-252, and MRC\,2048-262. For all three sources, ALMA detects a $J$\,$>$\,1 transition of CO along the radio axis but beyond the outer edge of the radio source, near the location of the previous ATCA detections of CO(1-0) (Paper I).\\
\vspace{-2mm}\\
$\bullet$ {\bf \object{MRC\,0114-211}:} This target contains a small, Compact Steep Spectrum (CSS) radio source with a length of $\sim$6 kpc \citep{bre10,ran11}. The reservoir of CO(1-0) was detected with the ATCA at 4.5$\sigma$ significance, and found along the radio jet at a distance of 34 kpc from the center of the galaxy (Paper I). The estimated molecular gas mass is $M_{\rm H_2}$\,=\,(4.5\,$\pm$\,0.9)\,$\cdot$\,$\alpha_{\rm CO}$\,$\times$\,10$^{10}$ $M_{\odot}$, with $\alpha_{\rm CO}$\,=\,$M_{\rm H2}$/$L^{\prime}_{\rm CO}$ the CO conversion factor that translates the CO luminosity ($L^{\prime}_{\rm CO}$) into $M_{\rm H2}$ \citep{bol13}, and which typically ranges from 0.8 for Ultra-Luminous Infrared Galaxies \citep{dow98} to 3.6 for high-$z$ starforming galaxies \citep{dad10,gen10}. Despite this large molecular gas mass, the region of the CO(1-0) emission is devoid of any emission in deep IRAC 4.5$\mu$m imaging (Fig. \ref{fig:mrc0114}).

Our ALMA data show a CO(2-1) detection roughly 0.8 arcsec beyond the radio source, in between the bright radio hot-spot and the previous CO(1-0) detection. The redshift of the CO(2-1) detection, $z$\,=\,1.414\,$\pm$\,0.001, is in good agreement with that derived from the optical [OIII] and H$\alpha$ lines \citep{nes17}, but shifted by $\sim$1500 \kms\ with respect to the peak of the CO(1-0) detection. The molecular gas mass associated with this CO(2-1) detection is $M_{\rm H_2(comp)}$ $<$ 1.8\,$\times$\,10$^{10}$\,$\cdot$\,$\alpha_{\rm CO}$ $M_{\odot}$ \citep[see][]{sol05}, where $\alpha_{\rm CO}$ is the CO-to-H$_{2}$ conversion factor \citep{bol13}. This estimate is based on the limit of r$_{\rm 2-1/1-0}$ given in Table \ref{tab:detect}.\\
\vspace{-2mm}\\
$\bullet$ {\bf \object{MRC\,0156-252}:} This target is associated with an extended radio source having a total linear size of $\sim$70 kpc and a spectacular 90$^{\circ}$ bend at its bright outer edge \citep{car97}. Luminous Ly$\alpha$ emission was found by \citet{pen01} to stretch along the radio axis, peaking at the location of the bright end of the radio source. The CO(1-0) emission observed with the ATCA at 5$\sigma$ significance is found beyond the radio source, while a region with diffuse X-ray emission is seen to the north of it \citep{ove05}. The CO(1-0) emission has a total mass of $M_{\rm H_2}$ $\sim$ (7.4\,$\pm$\,1.5)\,$\times$\,10$^{10}$\,$\cdot$\,$\alpha_{\rm CO}$ $M_{\odot}$ (Paper\,I) and is resolved into two components, which are separated by approximately 3 arcsec and 900 km\,s$^{-1}$ (regions B1 and B2 in Fig. \ref{fig:mrc0156}).

Our ALMA data show a clear CO(3-2) detection at a redshift of $z$\,=\,2.025\,$\pm$\,0.001 marked as region A in Fig.\,\ref{fig:mrc0156}, which is located close to the CO(1-0) component in region B1. This ALMA detection is co-spatial with a faint infra-red source in deep IRAC 4.5$\mu$m imaging (Fig.\,\ref{fig:mrc0156}), and therefore likely represents the molecular interstellar medium (ISM) in a companion galaxy. This source was identified by \citet{gal13} to be a red galaxy that is very faint in the optical, hence was not targeted by them for spectroscopic follow-up. Our ALMA data confirm that it is at about the same redshift as MRC\,0156-252.

Component B2, which was detected only in CO(1-0), is located in the region between the radio host galaxy and the companion in region A. The CO(1-0) peaks south-west of another, much bluer companion galaxy, which is found at the approximate location of the radio hot-spot \citep{gal13}. This companion galaxy is not visible in Fig. \ref{fig:mrc0156}. The redshift of this second companion is $z$\,=\,2.0171\,$\pm$\,0.0004 as derived from He\,II, which differs by $\sim$300 km\,s$^{-1}$ from the CO(1-0) velocity of the emission-line peak of component B2 (Fig.\,\ref{fig:mrc0156}). The CO(1-0) emission in region B2 may originate from this companion, or be a CGM component.

We also detect the host galaxy of MRC\,0156-252 in CO(3-2) at $z$\,=\,2.024\,$\pm$\,0.001, which is close to the redshift derived from He\,II of $z$\,=\,2.0256\,$\pm$\,0.0002 \citep{gal13}. There is a slight offset between the radio core and the peak of the central CO(3-2) emission, which lies $\sim$1 arcsec to the east, along the direction of the counter-jet. However, we cannot confirm that this offset is real, due to the uncertainties in the astrometry of the ALMA and radio continuum images, in particular because self-calibration of the VLA data \citep{car97} could have introduced an additional astrometric uncertainty of about a synthesized VLA beam.

The molecular gas masses associated with the CO(3-2) detections at the companion and the host galaxy are M$_{\rm H_2(comp)}$ $\sim$ (3.9\,$\pm$\,0.5)\,$\times$\,10$^{10}$\,$\cdot$\,$\alpha_{\rm CO}$ M$_{\odot}$ and M$_{\rm H_2(host)}$ $<$\,3.9 $\times$\,10$^{10}$\,$\cdot$\,$\alpha_{\rm CO}$ M$_{\odot}$, respectively \cite[see][]{sol05}, where $\alpha_{\rm CO}$ is the CO-to-H$_{2}$ conversion factor \citep{bol13}. These estimates are based on the values of r$_{\rm 3-2/1-0}$ given in Table \ref{tab:detect}.\\
\vspace{-2mm}\\
$\bullet$ {\bf \object{MRC\,2048-272}:} The CO(1-0) emission detected with the ATCA shows a profile with two peaks, one centred around v\,=\,0 \kms\ and the other around v\,=\,2500 \kms\ in Fig. \ref{fig:mrc2048}. The CO(1-0) emission covers a total velocity range of $\sim$3500 \kms, centred on the redshift derived from Ly$\alpha$ \citep{ven07}. The two spectral components were each detected at 3.5$\sigma$ significance, resulting in a combined 5$\sigma$ CO(1-0) detection of $M_{\rm H_2}$ $\sim$ (6.9\,$\pm$\,1.5)\,$\times$\,10$^{10}$\,$\cdot$\,$\alpha_{\rm CO}$ $M_{\odot}$ with ATCA (Paper\,I). The two components are co-spatial and located in the CGM, at a location that is devoid of any emission in IRAC 4.5$\mu$m imaging (Fig.\,\ref{fig:mrc2048}). 

Our ALMA results show a CO(3-2) detection associated only with one of the two spectral components of CO(1-0), at $z_{\rm CO(3-2)}$\,=\,2.050\,$\pm$\,0.001. The CO(3-2) signal is detected at a 4$\sigma$ level. The redshift and FWHM of the CO(3-2) emission match those of the blue spectral component of CO(1-0). This blue spectral CO(1-0) component contains $M_{\rm H_2}$ $\sim$ (4.5\,$\pm$\,1.1)\,$\times$\,10$^{10}$\,$\cdot$\,$\alpha_{\rm CO}$ $M_{\odot}$ (Paper\,I).\\
\vspace{-2mm}\\
Tables \ref{tab:detect} and \ref{tab:nondetect} further summarize the details of our ALMA results.

\begin{figure*}
\centering
\includegraphics[width=\textwidth]{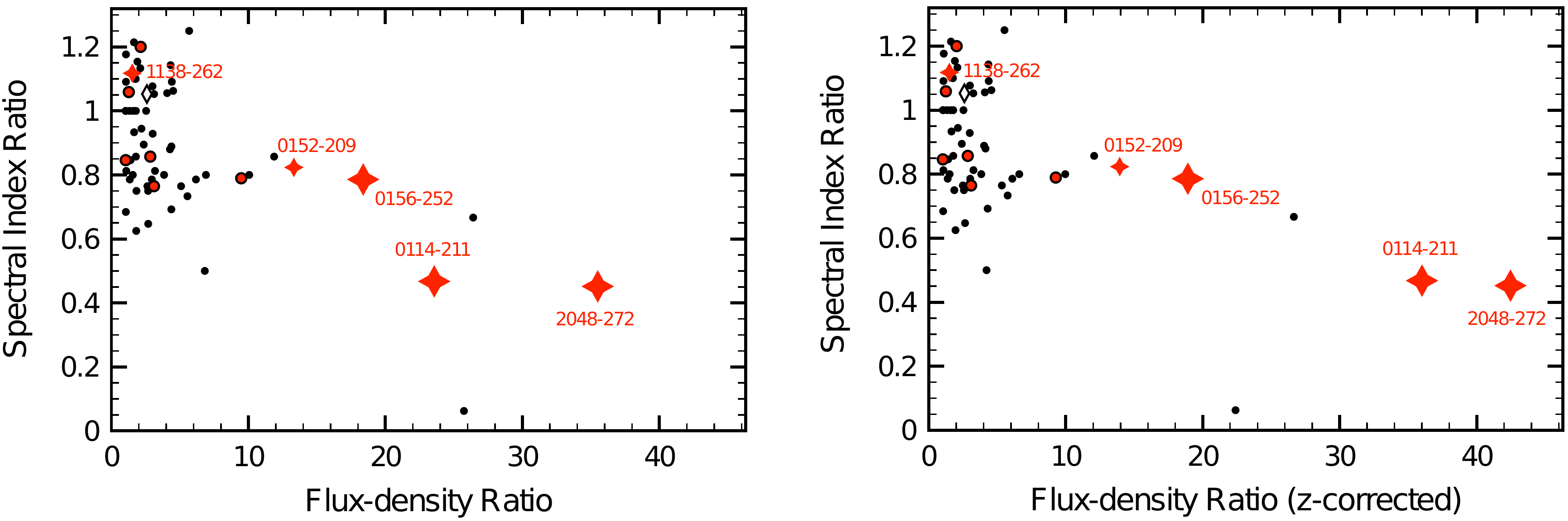}
\caption{Left: Plot showing on the horizontal axis the measured 4.7 GHz flux-density ratio between the lobe and counter lobe of high-$z$ radio galaxies. The vertical axis show the corresponding ratio of the 4.7$-$8.2 GHz spectral index of main and counter lobe. Higher lobe ratios indicate a larger difference in flux density between the main lobe and counter lobe, while progressively lower spectral-index ratios indicate that the brighter main lobe has a progressively shallower radio spectral index compared to the counter lobe. Red symbols show high-$z$ radio galaxies studied in CO(1-0) with ATCA (Paper I), with red stars the five CO(1-0) detections and small red dots the CO(1-0) non-detections. The large red stars are the three sources discussed in Sect. \ref{sec:almaresults} of this paper. The black dots are high-$z$ radio galaxies from the literature, acquired from \citet{car97}, \citet{pen00}, and \citet{bre10}. Measurements from the continuum imaging by \citet{bre10} are summarized in Appendix \ref{sec:continuum}. The black open diamond is TXS\,0828+193 (Sect.\,\ref{sec:boosting}). Right: same as left panel, but the flux densities and corresponding flux-density ratios are corrected for redshift, using the spectral index, to reflect a rest-frequency of $\nu_{\rm rest}$\,=\,16.0 GHz, which corresponds to $\nu_{\rm obs}$\,=\,4.7 GHz for a source at the median redshift of $z$\,=\,2.4 measured for the full literature sample. The spectral indices are not affected by this $z$-correction. 
}
\label{fig:ratios}
\end{figure*}

\subsection{Results on jet-CO alignment}
\label{sec:boosting}

The main observational result from this paper is that the ALMA data reveal reservoirs of cold molecular gas beyond the bright radio jet in three high-$z$ radio galaxies. This supports the previous ATCA detections of CO(1-0) in the environments, even though there are spatial and kinematical differences between the ALMA and ATCA detections.

In Fig. \ref{fig:ratios} we investigate how the presence of cold molecular gas in the environment relates to the properties of the radio source. For this analysis, we do not differentiate whether the molecular gas originates from a galaxy or the CGM. Our analysis is based on large samples of previously published 4.7 and 8.2 GHz radio synchrotron data of high-$z$ radio galaxies, which together have a median redshift of $z$\,=\,2.4 \citep{car97,pen00,bre10}. We compare the main lobe with the fainter counter lobe, by plotting the ratios of their flux densities and spectral indices in Fig. \ref{fig:ratios}. With the term `lobe' we mean the dominant hot-spot region at the resolution of the VLA data, given that only these values are available for the literature samples. 

Fig. \ref{fig:ratios} (left) shows the empirical result that MRC\,0114-211, MRC\,0156-252, and MRC\,2048-262 have a bright main lobe relative to the counter-lobe, with flux-density ratios of $S_{\rm main}$/$S_{\rm counter}$\,$\sim$24, 18, and 36 at 4.7\,GHz, respectively. These values are high compared to other high-$z$ radio galaxies from the literature shown in Fig.\,\ref{fig:ratios}, which show a median value of  $S_{\rm main}$/$S_{\rm counter}$\,$\sim$\,2.7. This suggests that the presence of cold molecular gas beyond the main lobe is directly related to the brightness of the synchrotron emission in these three systems.

One worry could be that the radio galaxies in Fig. \ref{fig:ratios} (left) have different redshifts, and therefore the measured 4.8 and 8.2 GHz flux densities correspond to a large range in rest frequencies. We can correct for this by adjusting the observed 4.7 GHz flux densities to values expected if the sources would all be at the median redshift of z\,=\,2.4 measured for the literature sample. For this, we use the spectral index $\alpha$ measured from the 4.8 and 8.2 GHz flux densities, where $S_{\nu}$\,=$\nu^{\alpha}$ (with $S_{\nu}$ the flux density). Using $\alpha$, we calculate the rest-frequency $\nu_{\rm rest}$\,=\,16 GHz, which corresponds to $\nu_{\rm obs}$\,=\,4.7 GHz at $z$\,=\,2.4. As shown in Fig.\,\ref{fig:ratios} (right), this results in an even stronger correlation.

To further investigate our results statistically, we consider that MRC\,0114-211, MRC\,0156-252, and MRC\,2048-262 were part of a representative sample of 13 high-$z$ radio galaxies that were studied in CO(1-0) with ATCA (Paper I). Besides MRC\,0114-211, MRC\,0156-252, and MRC\,2048-262, only two other high-$z$ radio galaxies from Paper I were detected in CO(1-0), namely MRC 0152-209 (a.k.a. Dragonfly galaxy) and MRC 1138-2623 (a.k.a. Spiderweb galaxy). Both the Dragonfly and Spiderweb radio galaxies also show extended CO(1-0) that stretches out to scales of 30-35 kpc distance from the host galaxy \citep{emo15a,emo15b,emo16,emo18}. Here we briefly described these two sources:
\begin{itemize}
\item{{\bf \object{MRC\,0152-209} (Dragonfly galaxy):} For the Dragonfly galaxy, \citet{leb23} conclude in a recent paper that the radio source brightens as it interacts with the molecular gas in the disk of a merging companion galaxy \citep[but see also][]{zho23}. This means that the Dragonfly galaxy follows the trend seen in Fig. \ref{fig:ratios}, with $S_{\rm main}$/$S_{\rm counter}$\,$\sim$13 at 4.7\,GHz. The CO(1-0) on large scales is likely tidal debris that stretches across a total extent of $\sim$60\,kpc \citep{emo15a}.}
\item{{\bf \object{MRC\,1138-262} (Spiderweb galaxy):} The Spiderweb galaxy \citep{mil06} shows no discrepancy in the 4.7\,GHz brightness ratio of the main and counter lobe ($S_{\rm main}$/$S_{\rm counter}$\,$\sim$\,1.5), but in this case the large-scale CO(1-0) emission was found in all directions across the halo. Moreover, while indications for alignments between the molecular gas and the radio source are more subtle in the Spiderweb, they are present, as evident from an increased brightness of CO(1-0) along the main jet \citep{emo16}, as well as the detection of water (H$_{2}$O) emission along the jet axis \citep{gul16b}. The large-scale CO(1-0) emission is likely halo gas that fuels in-situ star formation across $\sim$70\,kpc in the CGM \citep{hat08,emo16}.}
\end{itemize}
By including the Dragonfly and Spiderweb in our analysis, as well as the eight CO(1-0) non-detections from Paper I, we can perform a statistical analysis, which we will describe below. The only other high-$z$ radio galaxy mentioned in Sect. \ref{sec:intro} that has CO emission aligned with the radio source, and for which both 4.7 and 8.2 GHz observations of the radio source are available, is TXS\,0828+193 \citep{car97,nes09,fog21}. In the case of TXS\,0828+193, the CO emission was detected just outside the bright hot-spot of the main radio lobe \citep{nes09}, at the location of a companion galaxy \citep{fog21}. The main lobe on the side of this gas-rich companion is a factor $\sim$2.5 brighter than the counter-lobe (open diamond in Fig.\,\ref{fig:ratios}). However, because TXS\,0828+193 was not part of the representative MRC sample that was observed with ATCA in a uniform way, we do not take it into account in our statistical analysis.

Our statistical analysis is shown in Appendix \ref{sec:statistics}. We base our analysis on the conservative case where we do not correct the flux-density ratio values for redshift (i.e., left panel of Fig. \ref{fig:ratios}). When comparing the brightness ratios of the five CO-detected radio sources from Paper I (which all show large-scale CO emission) with those of the much larger literature sample of high-$z$ radio galaxies, a Kolmogorov-Smirnov (KS) test reveals that the probability that both samples are drawn from the same distribution is only $\sim$0.5$\%$ (see Appendix \ref{sec:statistics} for details). Therefore, the correlation in Fig. \ref{fig:ratios} between the presence of circumgalactic molecular gas beyond the radio source and an increased brightness-ratio between main lobe and counter-lobe, is statistically inconsistent with random sampling.

The correlation shown in Fig.\,\ref{fig:ratios} indicates that the increased brightness in the radio flux density of the main lobe is a direct result of the presence of cold molecular gas at or near the location of the radio hot-spot. The lobe brightening cannot be explained by Doppler boosting, which is the effect where the approaching radio jet increases its apparent brightness due to relativistic effects, because this only depends on the inclination angle of the jet and not on environmental conditions. Moreover, the 4.7 and 8.2 GHz measurements of the `lobe' predominantly reflect the hot-spot regions (Sect.\,\ref{sec:boosting}), which are the working surface where the lobe interacts with the CGM at sub-relativistic speeds. Although some Doppler boosting likely still occurs for the main hot-spot region, this typically results in a brightening of the flux density by a factor of at most a few \citep{kom96}, i.e., less than what we observe for MRC\,0114-211, MRC\,0156-252, MRC\,2048-272, and also MRC\,0152-209 (Dragonfly galaxy; \citealt{leb23}).

There is also a tentative indication from Fig. \ref{fig:ratios} that the 4.7-8.2 GHz spectral index is less steep within the bright main lobe among the three high-$z$ radio galaxies that we study in this paper. However, based on the same KS test as for the flux-density ratios, the difference in spectral-index values between the radio galaxies with detections of circumgalactic molecular gas and the large sample of high-z radio galaxies is not statistically significant. 

In summary, our results suggest that in the presence of molecular gas beyond the radio source, the radio synchrotron emission brightens. We will further discuss this result in the next Section.

\section{Discussion}
\label{sec:discussion}

The alignments that we observe between the radio sources and molecular gas reservoirs, as well as the increased flux density of the main lobe compare to the counter-lobe in the presence of these gas reservoirs, strongly suggest that jet-cloud interactions take place in the circumgalactic environment of our ALMA targets, even at distances out to $\sim$50 kpc from the radio host galaxy. \citet{eal92} suggested that such alignments can be explained by radio sources expanding into asymmetric gas distributions, and experiencing an increase in synchrotron luminosity that is on average largest in the direction where the gas density is highest. In addition, \citet{wes94} showed that radio sources tend to align with the major axis of the mass distribution on large scales. Therefore, these earlier theories agree with our observations. In this Section, we will explore the cause-and-effect that may lead to our observed alignments and lobe brightening.

\subsection{Jet-cloud interactions: synchrotron brightening}
\label{sec:interaction}

The brightening of the radio lobe in the presence of cold molecular gas suggests that the radio lobe is confined as it propagates into the molecular gas reservoir. As energy and momentum are being exchanged between the gas and the lobes, the gas experiences an increased velocity dispersion \citep[e.g.,][]{man21,mee22}. The fact that the CO(1-0) profiles of MRC\,0114-211, MRC\,0156-252, and MRC\,2048-262 are fairly broad, with a full width at zero intensity (FWZI) ranging from 1100 $-$ 3600 \kms, agrees with this and suggests that the molecular gas is turbulent. It is likely that compression and possible tangling of the magnetic fields, combined with the increased density of particles, boosts the radio synchrotron luminosity \citep[e.g.,][]{gop91,mor11}, although \citet{and22} show that jet-gas interactions also occur along radio sources with well-ordered, coherent magnetic fields that magnetize their surrounding environment. Among our sample, there is also a tentative indication that the spectral index of the main lobe becomes less steep compared to the counter lobe. If confirmed, this may indicate that shocks re-accelerate electrons in the main lobe, or that particle replenishment is still ongoing and therefore synchrotron losses due to spectral ageing are less severe in the lobe that is confined by interaction with the molecular gas \citep[e.g.,][]{blu00}.

The bulk of the molecular gas reservoirs, as identified by the peak of the CO emission, appear to be located well beyond the radio hot-spot. For MRC\,0156-252 and MRC\,2048-272, the distance between the hot-spot and the peak of the ALMA detected CO(3-2) emission appears to be $\sim$25 and $\sim$40 kpc, respectively. However, even though the working surface of the radio jet is thought to be the brightest there where it encounters the densest medium \citep[e.g.,][]{bar96}, the molecular gas reservoirs likely cover a large (tens of kpc) scale, and the interaction between the radio source and the molecular gas occurs at the boundary of the gas reservoir. The gas density even at the edge of the molecular reservoirs is likely high enough for this, considering that jet-gas interactions are frequently observed to be associated with much less denser gas, such as warm gas traced in Ly$\alpha$, H$\alpha$, and \oiii\ \citep[e.g.,][]{vil03,hum06,nes17}.

Under this scenario that jet-cloud interactions brighten the radio synchrotron emission, our observed alignments between radio sources and CO-emitting gas reservoirs could in part be due to intrinsically fainter radio sources being pushes into the flux-selected samples of high-$z$ radio galaxies \citep[see also][]{eal92}. This scenario was also suggested in the recent study of the Dragonfly galaxy (MRC\,0152-209) by \citet{leb23}. The scenario where jet-cloud interactions convert jet power into radio luminosity also agrees with studies of mostly ionized and neutral gas around low-$z$ radio sources \citep[e.g.,][]{bre85,fos98,vil99,vil17,tad00,mor02,mor11,mur20} and has been proposed to explain the relatively large fraction of young and compact radio sources seen among starbursting radio galaxies at low-$z$ \citep{tad11}.

\subsubsection{Jet-induced enrichment and cooling?}

Fig. \ref{fig:ratios} would be more difficult to explain if the molecular gas reservoirs would be merely a manifestation of the effects of the radio source, such as jet-induced enrichment and cooling, because this would likely be related to the {\sl total} radio power rather than the flux-density ratio of the lobe and counter-lobe. Nevertheless, in general, the most robust detections of large-scale (10s-100 kpc) molecular gas in circumgalactic environments are predominantly found to be associated with radio galaxies and radio-loud quasi-stellar objects (QSOs), while radio-quiet QSOs show less extended molecular gas reservoirs \citep{li23,jon23}. Therefore, it is worth exploring the effect that the radio jets may have on surrounding gas reservoirs. Radio-AGN activity is a recurrent phenomenon, so previous jet activity may also have contributed to creating the CO-emitting reservoirs.

Paper I described in detail that the effect of the radio jet on the circumgalactic medium may include chemical enrichment and gas cooling \citep[see also][]{kla04}. For nearby brightest clusters galaxies, \citet{kir09} and \citet{kir11} found that the radio source can drive chemical enrichment along the radio axis, typically over multiple episodes of radio-source activity. At high redshifts, such enrichment processes likely contribute to the near-solar metallicities observed across extended emission-line regions of high-$z$ radio galaxies and quasars \citep[e.g.,][]{ver01,hum08,pro09}. These processes could deposit the carbon and oxygen elements needed to form the large reservoirs of CO-emitting gas. 

In addition, the propagating radio jet likely compresses and cools the enriched CGM through shocks \citep[e.g.,][]{mel02,sut03,fra04,gai12,fra17,man21}. This process was suggested to occur in the Spiderweb radio galaxy (MRC\,1138-262), where both enhancements in CO(1-0) and H$_{2}$O emission were found along the radio jet, likely as a result of gas cooling behind slow shocks that propagate through the dense, multiphase gas \citep{gul16b,emo16}. Such gas cooling may result in jet-triggered star formation; see, for example, the case of Minkowski's Object \citep{cro06,sal15} 4C\,41.17 \citep{dey97,nes20}, and possible other high-$z$ radio galaxies \citep{kla04}. This could explain alignments found between high-$z$ radio jets and large-scale UV rest-frame continuum and submillimeter emission \citep{cha87,mcc87,beg89,you89,ree89,bic00,ste03,ivi12}, or the enhanced rates of star formation associated with radio-loud active galactic nuclei (AGN) \citep{zin13}.\\

\noindent In summary, while our CO results are in agreement with literature work indicating that radio sources may enrich or cool gas in their environment, our work (Fig.\,\ref{fig:ratios}) predominantly shows that the cold gas directly affects the radio source, enhancing its brightness due to jet-cloud interactions.

\subsection{Gas excitation: ISM or CGM?}

A detailed analysis of the physical properties of the molecular gas in the circumgalactic environment of MRC\,0114-211, MRC\,0156-252, and MRC\,2048-262 is difficult, given the low signal-to-noise and limited resolution of our ALMA and ATCA detections. Nevertheless, we here provide a first-order analysis of the properties of the molecular gas reservoirs, to obtain some insight into their nature.

For MRC\,0114-211, the ALMA detection of CO(2-1) (region A) peaks at roughly 10 kpc from the center of the radio galaxy and very close to the radio hot-spot. For MRC\,0156-252, the ALMA detection (region A) coincides with a detection in the IRAC Band 2 image, and is thus likely associated with a companion galaxy. For both cases, the line ratios of r$_{2-1/1-0}$\,$>$\,0.34 and r$_{3-2/1-0}$\,$\sim$\,0.30 (Table \ref{tab:detect}) are consistent with excitation conditions found in the ISM of high-z galaxies \citep[e.g.,][]{dan09,ivi11,bot13,ara14,dad15}. On the other hand, the line ratios are significantly lower for the CO(1-0) detections further out from the host or any companion galaxy, with r$_{2-1/1-0}$\,$<$\,0.13 for region B in MRC\,0114-211, r$_{3-2/1-0}$\,$<$\,0.22 for region B2 in MRC\,0156-252, and r$_{3-2/1-0}$\,$\lesssim$\,0.16 in MRC\,2048-262 (Table \ref{tab:nondetect}). Despite the uncertainty associated with this analysis, our results suggests that while part of the CO emission in MRC\,0114-211 and MRC\,0156-252 is likely associated with the ISM of either the host or a companion galaxy, at least some fraction of the CO(1-0) emission in these two systems, and possibly all of the CO emission in the third system MRC\,2048-272, appears to be part of the CGM. This molecular gas in the CGM has a low excitation, but it is not clear from our data whether it is diffuse or clumpy.

For future work, our results on the gas excitation strongly suggest that observations of the CO ground-transition that are sensitive to detecting low surface-brightness emission are critical for recovering widespread cold molecular gas reservoirs. ALMA Band 1 will soon allow observations of CO(1-0) out to redshifts of $z$\,$\la$\,2.3 ($\nu_{\rm obs}$\,$\ge$\,35 GHz). In terms of surface-brightness sensitivity, ALMA's extremely compact array configurations are an improvement over the VLA in D-configuration, whose Ka and Q band receivers operate in the same frequency regime. On longer time-scales, the Next-Generation VLA (ngVLA) will be able to target CO(1-0) at almost any redshift \citep{mur18,dec18}. In particular when optimized for surface-brightness sensitivity with a densely packed configuration of antennas in the central km-scale region \citep{car21}, the core of the ngVLA will be a critical complement to ALMA for tracing low-$J$ CO in the CGM at the highest redshifts \citep{emo18b}. Complementary ALMA observations of the high-$J$ CO transitions at higher frequencies can easily over-resolve or resolve out emission from widespread gas reservoirs. A large millimeter single-dish telescope, such as the Atacama Large Aperture Submillimeter Telescope (AtLAST; \citealt{ber18,cic19,kla20}), can recover all the flux of the high-$J$ CO lines.

\section{Conclusions}
\label{sec:conclusions}

We presented the detection of CO(2-1) or CO(3-2) emission in the environments of three high-$z$ radio galaxies, MRC\,0114-211 ($z$\,=1.41), MRC\,0156-252 ($z$\,=2.02), and MRC\,2048-272 ($z$\,=2.05). Our ALMA data support previous results from a CO(1-0) survey performed with ATCA, which detected CO(1-0) emission from cold molecular gas in the halo environments of these sources. We derive the following conclusions:
\begin{itemize}
    \item The CO is found along the radio axis but beyond the main radio lobe. This confirms that the `alignment effect', which is often observed for different constituents of high-$z$ radio galaxies, also applies to cold molecular gas on large scales in these sources.

    \item For MRC\,0114-211 and MRC\,0156-252, part of the observed molecular gas is likely ISM from the host or a companion galaxy, respectively. However, in other parts of the molecular reservoirs of these two sources, as well as for the case of MRC\,2048-272, the CO(2-1) or CO(3-2) emission does not appear to be associated with any galaxy and is faint compared to CO(1-0), with r$_{\rm 3-2/1-0}$ and r$_{\rm 2-1/1-0}$ values $\la$0.2. This suggests that we may also be tracing molecular gas with low excitation in the CGM. 

    \item The presence of cold molecular gas in the environment of our sources (whether from the ISM or CGM) correlates with an increase in brightness of the main radio lobe. This is derived from high ratios in the 4.7\,GHz flux density of the main lobe compared to the counter-lobe, with $S_{\rm main}$/$S_{\rm counter}$\,$\sim$\,18$-$36 for our three sources, compared to a median value of $S_{\rm main}$/$S_{\rm counter}$\,$\sim$\,2.7 for a large sample of high-$z$ radio galaxies from the literature. Based on a Kolmogorov-Smirnov test, we show that, at a statistical significance of 99.5$\%$, there is a difference in this ratio of the lobe brightness between high-$z$ radio galaxies with CO detected in the environment and the larger population of high-$z$ radio galaxies. This suggests that the cold gas affects the properties of the radio source, and that the alignment effect may in part be caused by intrinsically fainter radio sources which brightness is enhanced enough to enter the flux-selected samples of high-$z$ radio galaxies \citep[see also][]{leb23}. The radio source may also affect the molecular gas reservoir though enrichment and cooling, but this is more speculative from our data.
    
\end{itemize}

Our work confirms the importance of studying the interplay between radio sources and cold gas in the circumgalactic environments of high-$z$ galaxies. To understand how these results compare to the general population of galaxies in the Early Universe, future millimeter observations with compact interferometric arrays that are sensitive to recovering low-surface-brightness emission of low-$J$ CO are critical for studying the total budget of cold gas across the Universe.

\acknowledgments
We thank Donald Terndrup for useful discussions during the preparation of this paper. We also thank the referee for valuable feedback. This paper makes use of the following ALMA data: ADS/JAO.ALMA$\#$2015.1.00897.S. ALMA is a partnership of ESO (representing its member states), NSF (USA) and NINS (Japan), together with NRC (Canada), MOST and ASIAA (Taiwan), and KASI (Republic of Korea), in cooperation with the Republic of Chile. The Joint ALMA Observatory is operated by ESO, AUI/NRAO and NAOJ. The National Radio Astronomy Observatory is a facility of the National Science Foundation operated under cooperative agreement by Associated Universities, Inc. The Australia Telescope Compact Array is part of the Australia Telescope National Facility (https://ror.org/05qajvd42) which is funded by the Australian Government for operation as a National Facility managed by CSIRO. Based on observations with the NASA/ESA Hubble Space Telescope obtained from the Data Archive at the Space Telescope Science Institute, which is operated by the Association of Universities for Research in Astronomy, Incorporated, under NASA contract NAS5-26555. Support for program numbers HST-AR-16123.001-A and HST-GO-16891.002-A was provided through a grant from the STScI under National Aeronautics and Space Administration (NASA) contract NAS5-26555. M.V.M. acknowledges support from grant PID2021-124665NB-I00 by the Spanish Ministry of Science and Innovation (MCIN) / State Agency of Research (AEI) / 10.13039/501100011033 and by the European Regional Development Fund (ERDF) “A way of making Europe”.

\vspace{5mm}
\facilities{ALMA, ATCA, VLA, Spitzer, HST}


\software{CASA \citep{casa22}}



\clearpage

\appendix

\section{CO-bright galaxy in the field of MRC\,0114-211}
\label{sec:companion}. 

\begin{figure*}[b!]
\centering
\includegraphics[width=0.8\textwidth]{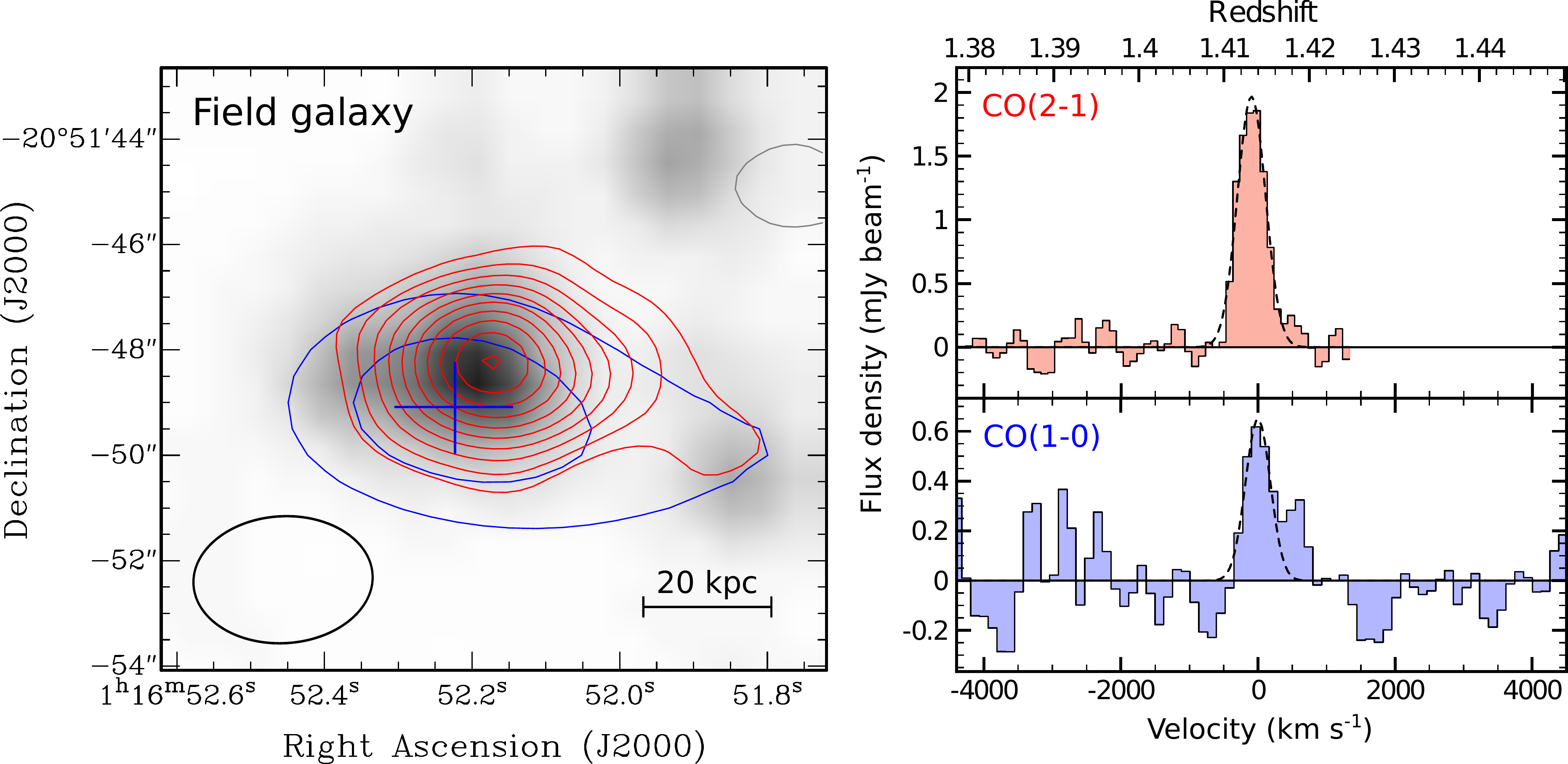}
\caption{Molecular gas in the CO-bright field galaxy north-west of MRC\,0114-211 ($z$\,=\,1.41). Left: IRAC 4.5$\mu$m image of the field galaxy \citep{gal12,wyl13}, with overlaid in red the contours of CO(2-1) observed with ALMA across the velocity range -468 to 332 \kms, and in blue contours of CO(1-0) observed with ATCA. CO contour levels start at 2$\sigma$ and increase with 1$\sigma$, with $\sigma$\,=\,0.085 Jy\,bm$^{-1}$\,$\times$\,\kms\ for CO(2-1) and $\sigma$\,=\,0.082 Jy\,bm$^{-1}$\,$\times$\,\kms\ for CO(1-0), after correcting for the primary beam response. Negative contours are shown in grey for CO(2-1), while for clarity they are omitted for CO(1-0). The blue cross marks the location of the CO(1-0) emission-line peak, while its size indicates the astrometric uncertainty. The astrometric uncertainty for the ALMA signal of CO(2-1) is not significant compared to CO(1-0) (see Table 1). The ALMA beam is indicated with the black ellipse in the bottom-left corner. Right: Spectra of CO(2-1) (top - red) and CO(1-0) (bottom - blue) associated with the field galaxy. The spectra were taken at the locations of the peaks of the CO emission, which for CO(1-0) is marked with a cross in the total intensity image shown in the left panel. Flux densities were corrected for the primary beam response. The dashed lines show single Gaussian fits to the CO profiles. The central velocity v\,=\,0 \kms\ is the same as in Fig.\,\ref{fig:mrc0114}, namely the redshift of MRC\,0114-211 derived from the CO(2-1) emission in region `A'.}
\label{fig:fieldgal}
\end{figure*}

A bright line-emitter was found in the field of MRC\,0114-211 at R.A.\,=\,01$^{\rm h}$\,16$^{\rm m}$\,52.17$^{\rm s}$; dec\,=\,-20$^{\circ}$\,51$^{\prime}$\,48.2$^{\prime\prime}$, which is a distance of $\sim$20$^{\prime\prime}$ ($\sim$170\,kpc) north-west of the radio galaxy. The line-emitter is detected in both CO(2-1) and CO(1-0), and has the same redshift as MRC\,0114-211 ($z$\,=\,1.414). This galaxy has a counterpart in the IRAC 4.5$\mu$m image of Fig.\,\ref{fig:mrc0114}.

The integrated CO(2-1) emission of this field galaxy is the strongest signal found among the ALMA data sets used in this paper, and is detected at an integrated signal-to-noise level of S/N\,$\approx$\,10. This means that the CO(2-1) emission in the ALMA data is bright enough to attempt a meaningful deconvolution and restoration (``cleaning") of the signal. We cleaned the signal in a circular aperture with a radius of 2$^{\prime\prime}$ centered on the peak of the CO(2-1) emission, down to a threshold of 2.5$\sigma$ in each channel. When processing and imaging the cleaned data in the same way as the data described in Sect.\,\ref{sec:observations}, the CO(2-1) peak flux-density of this field galaxy decreased by only 1.6$\%$ compared to the case of no deconvolution, while the peak flux-density of the CO(2-1) emission around MRC\,0114-211 decreased by only 2.4$\%$. These values are lower than the assumed 5$\%$ uncertainty in absolute flux calibration. Also, no signficant difference in the morphology of the signal is seen down to the 2$\sigma$ level shown in Fig.\,\ref{fig:mrc0114} when taking the same channel ranges for the integrated emission. Therefore, deconvolution does not affect the results presented in this paper, and at the low clean threshold that we had to adopt could include components that are dominated by noise. We therefore use the cleaned data cube only for the analysis of the field galaxy in this Appendix.

Figure \ref{fig:fieldgal} shows the CO(2-1) and CO(1-0) detection of this field galaxy. The data properties as the same as in Table \ref{tab:data}, with the difference that we applied a primary beam correction to recover accurate flux values this far from the center of the image. To derive CO intensities, we fit a Gaussian model to the CO lines. The resulting line properties for the ALMA CO(2-1) [ATCA (CO(1-0)) data] are: $v_{\rm CO}$\,=\,-100$\pm$\,10 [0\,$\pm$\,60] \kms; FWHM$_{\rm CO}$\,=\,480\,$\pm$\,30 [430\,$\pm$\,130] \kms; $I_{\rm CO}$\,=\,1.01\,$\pm$\,0.08 [0.30\,$\pm$\,0.08] \mjybm\,\kms. The total H$_{2}$ mass of the field galaxy derived from CO(1-0) is M$_{\rm H_2}$\,=\,(3.0\,$\pm$\,0.8)\,$\times$\,10$^{10}$ $\cdot$ $\alpha_{\rm CO}$ $M_{\odot}$ \citep{sol05}, and the ratio of the CO luminosities is r$_{2-1/1-0}$\,=\,0.84\,$\pm$\,0.23. This is consistent with values found for other high-$z$ galaxies \citep[e.g.,][]{dan09,ivi11,bot13,ara14,dad15}. There is a tentative indication that the CO emission may be extended towards the east, but this needs to be confirmed.

\section{Radio continuum properties of three ATCA sources}
\label{sec:continuum}

In this Appendix, we provide the continuum properties of three sources that are part of our ATCA sample, MRC\,0114-211, MRC\,0324-228, and MRC\,0350-279, which were all three imaged at 4.7 GHz and 8.2 GHz by \citet{bre10}. The flux densities and spectral indices are measured from the continuum images of \citet{bre10} and results are summarized in Table \ref{tab:continuum}.

\begin{deluxetable}{lcccc}[htb!]
  \tablecaption{Continuum properties three ATCA sources}
\tablecolumns{5}
\tablenum{4}
\tablewidth{0pt}
\tablehead{
\colhead{Source} & \multicolumn{2}{c}{Brightest hot-spot} & \multicolumn{2}{c}{Faintest hot-spot} \\
\colhead{} & \colhead{$I_{\rm 4.7\,GHz}$$^{\dagger}$} & \colhead{$\alpha_{\rm 4.7}^{8,2}$} & \colhead{$I_{\rm 4.7\,GHz}$$^{\dagger}$} & \colhead{$\alpha_{\rm 4.7}^{8,2}$} 
}
\startdata
MRC\,0114-211 & 1.06 & -1.1 & 0.045 & -2.3 \\
MRC\,0324-228 & 0.041 & -1.8 & 0.019 & -1.5 \\
MRC\,0350-279 & 0.033 & -1.8 & 0.026 & -1.7 \\
\enddata
\tablecomments{$^{*}$ Units for $I_{\rm 4.7\,GHz}$ are in Jy\,beam$^{-1}$.}
\label{tab:continuum}
\end{deluxetable}

\section{Statistical analysis}
\label{sec:statistics}

\begin{figure*}[h!]
\centering
\includegraphics[width=0.95\textwidth]{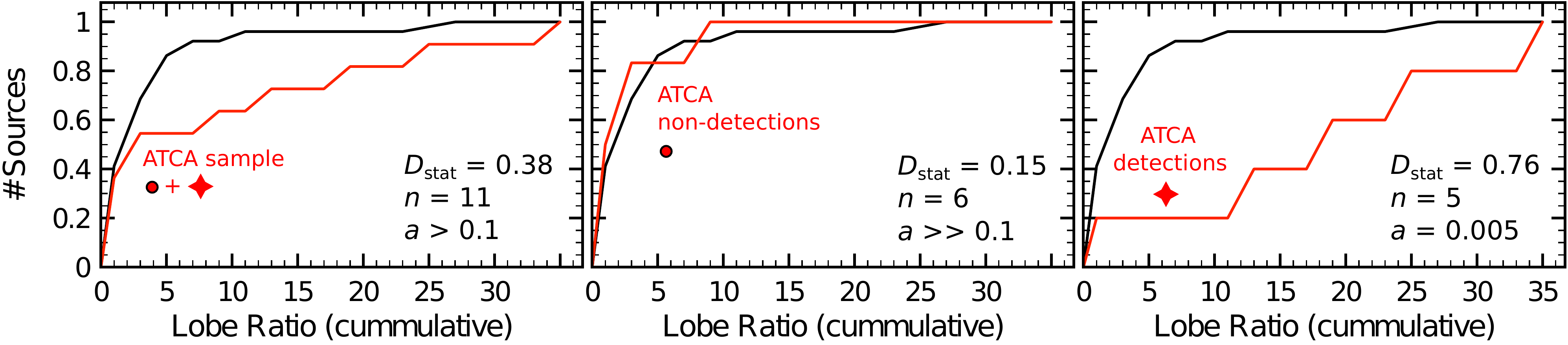}
\caption{Statistical analysis on the significance of high lobe-ratio values among the high-$z$ radio galaxies with extended CO emission, using a one-tailed Kolmogorov-Smirnov (KS) test for two samples of unequal size. {\sl Left (A):} Cummulative distribution of the 4.7 GHz lobe ratios for the full ATCA CO(1-0) sample of 11 sources (detections and non-detections) from Fig. \ref{fig:ratios} (red line), compared to the other 51 high-$z$ radio galaxies from the literature shown in Fig. \ref{fig:ratios} that were not targeted with ATCA (black line). {\sl Middle (B):} same as left panel, but only including the 6 ATCA CO(1-0) non-detections (red line). {\sl Right (C):} same as left panel, but only including the 5 ATCA CO(1-0) detections. In the legend, $D_{\rm stat}$ is the Kolmogorov-Smirnov test statistic, $n$ is the number of sources in the ATCA sample that are included in the analysis, and $a$ is the probability that the two samples are drawn from the same probability distribution. Only the sample of the ATCA CO detections (panel C) shows a statistically significant difference in lobe ratio with the literature sample of high-$z$ radio galaxies.} 
\label{fig:stats}
\end{figure*}

In this Appendix, we investigate whether there is a statistical significance to the observed correlation that the high-$z$ radio sources with extended CO emission also have a relatively bright main lobe compared to the counter lobe (Fig.\,\ref{fig:ratios} - left). For this, we use a one-tailed Kolmogorov-Smirnov (KS) test for two samples of unequal size. The first sample is our ATCA sample from Paper I. Out of the 13 MRC sources observed by ATCA, only 11 have information regarding their radio continuum properties available in \citet{car97}, \citet{pen00}, or \citet{bre10}. We divide these 11 MRC sources into a sub-sample of five detections and a sub-sample of 6 non-detection in CO(1-0). The non-detections from the ATCA sample have a 3$\sigma$ upper limit of $L^{\prime}_{\rm CO(1-0)}$ that is equal to or smaller than $L^{\prime}_{\rm CO(1-0)}$ of the five ATCA detections \citep{emo14}. Therefore, we do not expect a bias regarding the CO(1-0) sensitivity across the sample observed with ATCA. The second sample is the literature sample of high-$z$ radio galaxies shown in Fig. \ref{fig:ratios}, but excluding the MRC sources that we observed with ATCA. Fig. \ref{fig:stats} shows the results from our KS analysis. 

Panel A of Fig.\,\ref{fig:stats} shows that when comparing the literature sample with the full ATCA sample (detections and non-detections), the probability that both samples are drawn from the same distribution is just above 10$\%$. This means that there is no statistical difference between the two samples, which in turn means that there were no significant biases that influenced the selection of the ATCA sample in terms of the lobe-ratio values shown in Fig. \ref{fig:ratios}. Panel B shows that the same conclusion can be drawn when considering just the non-detections in the ATCA sample (red dots in Fig. \ref{fig:ratios}), with an even better correspondence with the literature sample. However, panel C shows that when considering only the CO detections in the ATCA sample (stars in Fig. \ref{fig:ratios} - left), the probability that this sample of 5 sources is drawn from the same distribution as the literature sample is only 0.5$\%$. This means that there is a statistical difference in the ratio of the lobe brightness between our sample of radio galaxies with CO detected in the environment and the larger population of high-$z$ radio galaxies. We note that this result is obtained by including MRC\,1138-262, which shows extended CO(1-0) emission in all directions across the halo \citep{emo16}. If we would only consider the four ATCA sources of interest for this paper, namely those sources with extended CO detected along the radio axis, then this would even further decrease the probability that their high ratio in lobe brightness is the result of random sampling.

\end{document}